\documentclass[twocolumn,showpacs,preprintnumbers]{revtex4}
\usepackage{amssymb}
\usepackage{amsmath}
\usepackage{graphicx}
\usepackage{dcolumn}
\usepackage{bm}

\begin{document}

\title{\textbf{BCS-BEC model of high-}$\mathbf{T}_{\mathbf{c}}$ \textbf{%
superconductivity in layered cuprates} \textbf{with unconventional pairing }}
\author{C. Villarreal$^{1}$ and M. de Llano$^{2}$}
\date{\today }
\affiliation{$^{1}$Instituto de F\'{\i}sica, Universidad Nacional Aut\'{o}noma de M\'{e}%
xico Ciudad Universitaria, 04510 M\'{e}xico, DF, MEXICO \\
$^{2}$Physics Department, University of Connecticut, Storrs, CT 06269 USA and%
\\
$^{\ast }$Instituto de Investigaciones en Materiales, Universidad Nacional
Aut\'{o}noma de M\'{e}xico Apdo.Postal 70-360, 04510 M\'{e}xico, DF, MEXICO}

\begin{abstract}
High-$T_{c}$ superconductivity in layered cuprates is described in a BCS-BEC
formalism with linearly-dispersive s- and d-wave Cooper pairs moving in
quasi-2D finite-width layers about the $CuO_{2}$ planes. This yields a
closed formula for $T_{c}$ determined by the layer width, the Debye
frequency, the pairing energy, and the in-plane penetration depth. The new
formula reasonably reproduces empirical values of superconducting $T_{c}$s
for seven different compounds among the $LSCO,$ $YBCO,$ $BSCCO$ and $TBCCO$\
layered cuprates.
\end{abstract}

\maketitle

%\preprint{APS/123-QED}

\section{Introduction}

It seems to be well-established that central to high-$T_{c}$
superconductivity (HTSC) is the layered two-dimensional (2D) structure of
copper oxides and that superconducting pairing occurs mainly on the $CuO_{2}$
planes. However, the precise nature of the pairing is still the subject of
intense research. Recent experiments based on angle-resolved photon emission
spectroscopy (ARPES) of underdoped cuprates suggest that bound fermion
Cooper pairs (CPs) form already at and below temperatures higher than the
critical transition temperature $T_{c}$ \cite%
{tanaka06,valla06,millis06,yang08}. Furthermore, ARPES studies of the
electron spectral function of optimally doped Bi2212 samples show that the
magnitude of the isotope effect correlates with the superconducting gap \cite%
{gweon04}, thus suggesting a role of lattice phonons in the superconducting
pairing. ARPES data also suggest that the energy gap $\Delta $ (a measure of
the energy needed to break a CP) displays an unconventional $d_{x^{2}-y^{2}}$
orbital pairing symmetry, with a functional dependence $\Delta =\Delta
_{0}\cos 2\theta $ where $\theta =\tan ^{-1}K_{y}/K_{x}$ is the angle
between the total or center-of-mass momentum (CMM) $\hbar K=(\hbar
K_{x},\hbar K_{y})$ of paired electrons in the $CuO_{2}$ plane and the $a$-
(or $x$-) axis while $\Delta _{0}$ is the value of the superconducting gap
at the antinode ($\theta =0,\pi /2$) \cite{maki94}. This behavior is also
apparent in studies based on electronic Raman scattering \cite{letacon06}
and in determinations of the in-plane magnetic penetration depth $\lambda
_{ab}$ \cite{panagopoulos98,broun07}.

Although the majority viewpoint in the high-$T_{c}$ community seems to argue
for such non-s-wave pairing symmetry there are compelling dissenting views,
particularly work within the past few years, by M\"{u}ller \cite{muller02},
Harshman \cite{harshman04}, Klemm \cite{klemm05}, and many others. In
particular, M\"{u}ller concludes that \textquotedblleft ...recent
experiments probing the surface and bulk of cuprate superconductors [show
that their] character is $d$ on the surface and substantially $s$ in the
bulk.\textquotedblright\ This conclusion has been bolstered by muon-spin
relaxation ($\mu $SR) experiments with YBCO reported and interpreted in Ref.%
\cite{harshman04}. Several authors \cite{pashitiskii97,fujita02} have
proposed that the doping process could modify the electron-phonon
interaction and the Fermi surface with a concomitant shift from $d$- to $s$%
-type coupling as doping increases. The strongest evidence for an s-wave
order parameter in a cuprate is reviewed in Ref.\cite{klemm05} where several 
$c$-axis twist experiments on BSCCO along with earlier $c$-axis tunneling
between BSCCO/Pb junctions are surveyed. Ref.\cite{arnold06} summarizes many
of the problems with the so-called \textquotedblleft
phase-sensitive\textquotedblright\ tests \cite{tseuiRMP00} in YBCO.
Additionally, predictions made in Ref.\cite{ichiokaPRB99} that a vortex in a 
$d$-wave superconductor would exhibit a measurable density of states in a
four-fold pattern emanating from the core have not been observed \cite%
{klemm09privcomm}\ in either YBCO or BSCCO. On the contrary, the vortex
cores appear consistent \cite{klemm09privcomm} with isotropic $s$-waves.

Replacing the characteristic phonon-exchange Debye temperature $\Theta
_{D}\equiv \hbar \omega _{D}/k_{B}$ of around $400K$ (with $k_{B}$ the
Boltzmann constant and $\omega _{D}$ the Debye frequency) by a
characteristic \textit{magnon}-exchange temperature of around $1000K$ can
lead to a simple model interaction such as the BCS one but associated with
spin-fluctuation-mediated pairing \cite{dahm09}.

The so-called \textquotedblleft Uemura plot\textquotedblright\ \cite%
{uemura04} of data from $\mu $SR, neutron and Raman scattering, and ARPES
measurements exhibit $T_{c}$ \textit{vs} Fermi temperatures $T_{F}\equiv
E_{F}/k_{B}$ where $E_{F}$ the Fermi energy and $k_{B}$ the Boltzmann
constant. Empirical $T_{c}$s\ of many cuprates straddle a straight line 
\textit{parallel} to the Uemura-plot line associated with the simple BEC
formula $T_{B}\simeq 3.31\hbar ^{2}n_{B}^{2/3}/m_{B}k_{B}\simeq 0.218T_{F}$
corresponding to an ideal gas of bosons of mass $m_{B}=2m^{\ast }$ and
number density $n_{B}=n_{s}/2$ where $m^{\ast }$ is the effective mass and $%
n_{s}$ the number density of individual charge carriers. The parallel line
is shifted down from $T_{B}$\ by a factor $4$-$5$. This has been judged \cite%
{Uemura06}\ as a \textquotedblleft fundamental importance of the BEC concept
in cuprates.\textquotedblright

Previous theoretical papers on the possible origin of HTSC \cite%
{fujita02,dellano98,FG,dellano07} proposed that the phenomenon might be
rooted in a 2D Bose-Einstein condensate (BEC) of CPs pre-existing above $%
T_{c}$ and coupled through a BCS-like phonon mechanism \cite{BCS},
originally taken as $s$-wave. As apparently first reported by Schrieffer 
\cite{schrieffer64}, the Cooper model interaction \cite{cooper56} leads to
an approximate \textit{linear} energy-vs-CMM relation $%
%TCIMACRO{\U{bd}}%
%BeginExpansion
{\frac12}%
%EndExpansion
v_{F}\hbar K$ for excited CPs propagating in the Fermi sea in 3D. This kind
of dispersion relation is not unique to the Cooper model interaction. For
example, an attractive interfermion delta potential \cite{Miyake} in 2D
(imagined regularized \cite{GT} to support a single bound state of binding
energy $B_{2}\geqslant 0$) leads \cite{PRB2000} (for a review see Ref.\cite%
{dellano07}) to the general CP dispersion relation $\mathcal{E}_{K}=\mathcal{%
E}_{0}+c_{1}\hbar K+\left[ 1-(2-16/\pi ^{2})E_{F}/B_{2}\right] \hbar
^{2}K^{2}/4m^{\ast }+O(K^{3})$, where $c_{1}=2v_{F}/\pi $ precisely as with
the Cooper model interaction \cite{PC98}. Hence, the leading-order linearity
is not induced by the particular interfermion interaction binding the CPs
but is a consequence of the Fermi sea with $v_{F}\neq 0$\ and in which a CP
by definition propagates. Only in the vacuum limit $v_{F}\rightarrow
0\Rightarrow E_{F}\equiv 
%TCIMACRO{\U{bd}}%
%BeginExpansion
{\frac12}%
%EndExpansion
m^{\ast }v_{F}^{2}\rightarrow 0$ does that general dispersion relation
reduce by inspection to the expected quadratic form $\mathcal{E}_{K}=%
\mathcal{E}_{0}+\hbar ^{2}K^{2}/4m^{\ast }$ for a composite object of mass $%
2m^{\ast }$. For either interelectron interaction model, the linear term is
a consequence of the presence of the Fermi sea. The formation of a BEC of
CPs in 2D does not violate Hohenberg's theorem \cite{hohenberg67} as this
holds only for quadratically-dispersive particles. The predicted 2D BEC
temperature is $T_{c}\propto \left( n^{2D}\right) ^{1/2}\propto \left(
\Theta _{D}T_{F}\right) ^{1/2}$ where $n^{2D}$ is the CP number per unit
area. This leads to values of $T_{c}$ that are too high compared with
empirical values. However, these schemes provide a correct description of
other relevant physical properties of HTSCs such as a short coherence
length, a type II magnetic behavior, and the temperature dependence of the
electronic heat capacity \cite{fujita02}. They also lead to excellent fits
of the condensate fraction curves for quasi-2D cuprates just below $T_{c}$ 
\cite{dellano06}, as well as for 3D and even quasi-1D SCs. To go beyond the
simple $s$-wave interaction, an $l$-wave formulation of BCS theory was
discussed by Schrieffer \cite{schrieffer64} himself and studied in
considerable detail by Anderson and Morel \cite{anderson61} in the
weak-coupling limit. This has been successfully employed \cite%
{hirschfeld93,maki94} to describe thermodynamic and transport properties of
high-$T_{c}$ cuprates. The $d$-wave extension in strong-coupling Eliashberg
theory is reported in Refs.\cite{carbotte}.

Here we develop a general $l$-wave BCS-type theory which is then applied in
a quasi-2D BEC picture with either $l=0$ or $l=2$ pairing symmetry. In \S\ %
II the $l$-wave BCS theory within the framework of the present model is
discussed. In \S\ III we study a quasi-2D BEC of linearly-dispersive, {%
massless-like} CPs and we evaluate the number density. In \S\ IV the areal
density $n^{2D}$ of charge carriers is estimated by calculating the magnetic
penetration depth arising from {the} CPs. In \S\ V an analytical expression
for the critical BEC temperature is derived, which is then applied in \S\ VI
for various superconducting materials including YBCO under different doping
levels. Discussion and conclusions are given in \S\ VII.

\section{BCS theory with $l$-wave pairs}

Some aspects of the $l$-wave BCS theory \cite{schrieffer64,anderson61,maki94}
relevant to our HTSC model follow. Consider a system of electron- (or hole-)
pairs formed via a two-fermion isotropic potential $V$ near the Fermi
surface and with kinetic energies $\epsilon _{k}\equiv \hbar
^{2}(k^{2}-k_{F}^{2})/2m^{\ast }$ (with $\hbar k_{F}$ the Fermi momentum)
taken relative to the Fermi energy. The Pauli principle prevents background
fermions in electron states just below (above) the Fermi surface from
participating in the interaction. In the absence of external forces each
pair propagates freely within a layer of finite width $\delta $ along the $z$
direction and infinite extent on the $x-y$ plane so that its total momentum $%
\hbar \mathbf{K}=(\hbar \mathbf{K}_{\parallel },\hbar K_{z})$ is a constant
of motion. By neglecting spin-dependent interactions the total spin $S$ is
conserved too and for a spin singlet $S=0$ configuration the Pauli principle
requires that the orbital wavefunction be of the form $\Psi (\mathbf{r}_{1},%
\mathbf{r}_{2})=\exp (i\mathbf{K}_{\parallel }\cdot \mathbf{R}_{\parallel
})\cos (K_{z}z)\Phi (\mathbf{r})$, where the relative coordinate $\mathbf{r}=%
\mathbf{r}_{1}-\mathbf{r}_{2}$, $\mathbf{R}_{\parallel }$ is the horizontal
projection of the CM coordinate $\mathbf{R}=(\mathbf{r}_{1}+\mathbf{r}_{2})/2
$, and $K_{z}=n\pi /\delta $ (with $n$ integer). The $z$-dependence of the
wavefunction ensures that the vertical flux of the electron (hole) pair
across the layer boundary is null. Since the relative-coordinate problem is
isotropic then $\Phi (\mathbf{{r})}$ is an eigenfunction of angular momentum
with quantum numbers $l=0,1,2,\cdots $. The total spin $S=0$ singlet
eigenstates of the system satisfy the Schr\"{o}dinger equation 
\begin{equation}
(H_{0}-V)\Psi (\mathbf{r}_{1},\mathbf{r}_{2})=\mathcal{E}\Psi (\mathbf{r}%
_{1},\mathbf{r}_{2})  \label{sch}
\end{equation}%
where $H_{0}$ is the free Hamiltonian $V$ the interaction potential and $%
\mathcal{E}$ the energy eigenvalue. For a given CMM wavevector $\mathbf{K}$,
we may expand the wave function as%
\begin{equation}
\Psi (\mathbf{r}_{1},\mathbf{r}_{2})=\exp (i\mathbf{K}_{\parallel }\cdot 
\mathbf{R}_{\parallel })\cos (K_{z}z)\sum_{\mathbf{k}}a_{\mathbf{k}}\exp (i%
\mathbf{k}\cdot \mathbf{r).}  \label{psi}
\end{equation}%
In momentum space (\ref{sch})\ thus becomes 
\begin{equation}
\left( \mathcal{E}-\epsilon _{\mathbf{k}+\mathbf{K}/2}-\epsilon _{\mathbf{k}-%
\mathbf{K}/2}\right) a_{\mathbf{k}}=\sum_{\mathbf{k}^{\prime }}V_{\mathbf{k}%
\mathbf{k}^{\prime }}a_{\mathbf{k}^{\prime }}  \label{sch1}
\end{equation}%
with $V_{\mathbf{k}\mathbf{k}^{\prime }}=$ $<\mathbf{k},-\mathbf{k}|V|%
\mathbf{k}^{\prime },-\mathbf{k}^{\prime }>$. Since the interaction
potential $V$\ depends only on $\mathbf{r}$ it admits the expansion 
\begin{equation}
V_{\mathbf{k}\mathbf{k}^{\prime }}=\sum_{l=0}^{\infty }\sum_{m=-l}^{l}V_{l}(|%
\mathbf{k}|,|\mathbf{k}^{\prime }|)Y_{l}^{m}(\Omega _{\mathbf{k}%
})Y_{l}^{-m}(\Omega _{\mathbf{k}^{\prime }}).  \label{pot}
\end{equation}%
For small coupling amplitudes $V_{l}(|\mathbf{k}|,|\mathbf{k}^{\prime }|)$
the contributions of different $l$ spherical harmonics $Y_{l}^{m}(\Omega _{%
\mathbf{k}})$ in (\ref{pot}) can with good accuracy be considered relatively
independent \cite{anderson61}. In that case (\ref{sch1}) yields an
analytical solution by assuming that $V_{l}$ is separable, i.e., $V_{l}(|%
\mathbf{k}|,|\mathbf{k}^{\prime }|)=V_{0}^{(l)}f_{k}^{l}f_{k^{\prime
}}^{l\ast }$ so that (\ref{sch1}) becomes%
\begin{equation}
\left( \mathcal{E}-\epsilon _{\mathbf{k}+\mathbf{K}/2}-\epsilon _{\mathbf{k}-%
\mathbf{K}/2}\right) a_{\mathbf{k}}=V_{0}^{(l)}f_{k}^{l}\sum_{k^{\prime
}}a_{k^{\prime }}f_{k^{\prime }}^{l\ast }  \label{sch2}
\end{equation}%
where $a_{\mathbf{k}}=a_{k}Y_{l}^{m}(\Omega _{\mathbf{k}})$. Eq.(\ref{sch2})
can be now rewritten as 
\begin{equation}
a_{k}^{(l)}=C^{(l)}\frac{V_{0}^{(l)}f_{k}^{l}}{\mathcal{E}%
_{K}^{(l)}-\epsilon _{\mathbf{k}+\mathbf{K}/2}-\epsilon _{\mathbf{k}-\mathbf{%
K}/2}}  \label{sch3}
\end{equation}%
where $\sum_{\mathbf{k}^{\prime }}a_{k^{\prime }}f_{k^{\prime }}^{l\ast
}\equiv C^{(l)}$ is a constant. One thus obtains a BCS-type integral
relation for a CP in the eigenstate characterized by $(l,m)$%
\begin{equation}
1=V_{0}^{(l)}\sum_{k}\frac{|f_{k}^{l}|^{2}}{\mathcal{E}_{K}^{(l)}-\epsilon _{%
\mathbf{k}+\mathbf{K}/2}-\epsilon _{\mathbf{k}-\mathbf{K}/2}}.  \label{BCS}
\end{equation}%
Following Schrieffer \cite{schrieffer64} we assume that the
angular-independent $l$ component of the generalized BCS model interaction (%
\ref{pot}) is given by 
\begin{equation}
V_{0}^{(l)}f_{k}^{l}f_{k^{\prime }}^{l\ast }=-V_{0}  \label{interaction}
\end{equation}%
with $V_{0}>0$ for CPs with relative momenta $(\mathbf{k},\mathbf{k}^{\prime
})$ lying in the neighborhood of the Fermi surface 
\begin{equation}
k_{F}<|\mathbf{k}+\mathbf{K}/2|,|\mathbf{k}-\mathbf{K}/2|<K_{max}
\label{slab}
\end{equation}%
and $V_{0}^{(l)}f_{k}^{l}f_{k^{\prime }}^{l\ast }=0$ otherwise. Here $%
K_{max}=\sqrt{k_{F}^{2}+k_{D}^{2}}$ with $k_{D}$ defined in terms of the
Debye energy via $\hbar \omega _{D}\equiv \hbar ^{2}k_{D}^{2}/2m^{\ast }$. A
straightforward analysis \cite{schrieffer64} reveals that (\ref{BCS}) yields
a bound state with energy $\mathcal{E}^{(l)}<0$ for arbitrarily weak
coupling so long as the potential is attractive in the region (\ref{slab})
in $k$-space. Then, a bosonic CP can form only if the tip of vector $\mathbf{%
k}$ lies within the intersection of the two spherical shells defined by (\ref%
{slab}) whose center-to-center separation is $K$; fermions with wave vectors
lying outside this overlap are unpairable \cite{dellano07}.

In the quasi-2D limit the fundamental expression (\ref{BCS}) can be
evaluated by substituting the summation over $\mathbf{k}$ by a 2D
integration. In addition, for small $\delta $ the only term in $K_{z}$ that
yields a finite contribution is $n=0$. By assuming a 2D cylindrical Fermi
surface we obtain 
\begin{eqnarray}
1 &=&\frac{V_{0}}{(2\pi )^{2}}\int_{0}^{2\pi }d\theta \int_{k_{1}}^{k_{2}}%
\frac{kdk}{|\mathcal{E}_{K}^{(l)}|+\epsilon _{\mathbf{k}+\mathbf{K}%
/2}+\epsilon _{\mathbf{k}-\mathbf{K}/2}} \\
&\simeq &\frac{m^{\ast }V_{0}}{4\pi }{\Big \langle}\ln {\Big \vert}\frac{|%
\mathcal{E}_{K}^{(l)}|+2\hbar \omega _{D}-v_{F}\hbar K\cos \theta }{|%
\mathcal{E}_{K}^{(l)}|+v_{F}\hbar K\cos \theta }{\Big \vert}{\Big \rangle}%
_{F}  \notag  \label{angular}
\end{eqnarray}%
where $k_{1}=k_{F}+(K/2)\cos \theta $, $k_{2}=k_{F}+k_{D}-(K/2)\cos \theta $%
, and the approximate equality in the second row holds up to terms of order $%
(k_{D}/k_{F})^{2}\equiv \Theta _{D}/T_{F}$. The angular brackets denote an
average over a 2D cylindrical Fermi surface $<...>_{F}\rightarrow (1/2\pi
)\int_{0}^{2\pi }d\theta $. The Fermi average then gives the energy spectrum
of excited CPs \cite{schrieffer64}: 
\begin{equation}
\mathcal{E}_{K}^{(l)}\simeq \mathcal{E}_{0}^{(l)}+c_{1}\hbar K+O(K^{2})
\label{dispersion}
\end{equation}%
where $c_{1}\equiv 2v_{F}/\pi $ in 2D and $\mathcal{E}_{0}^{(l)}$ is the
binding energy of the CP ground state ($\hbar K=0$) \cite{cooper56} 
\begin{equation}
\mathcal{E}_{0}^{(l)}=-2\hbar \omega _{D}/[\exp (2/\mathcal{N}%
_{0}V_{0}^{(l)})-1].  \label{epsilon0}
\end{equation}%
The dispersion relation (\ref{dispersion}) is linear in leading order rather
than quadratic as would be expected \textit{in vacuo}. As a consequence, all 
\textit{excited} CPs behave like a gas of free massless-like bosons with a
common group velocity $c_{1}=\hbar ^{-1}d\mathcal{E}_{K}^{(l)}/dK$, but a
variable energy determined by their CMM $\hbar K$. The dispersion relation (%
\ref{dispersion}) implies that in order for a CP to remain bound (i.e., $%
\mathcal{E}_{K}^{(l)}<0$) its maximum CMM wavenumber must not exceed the
value $|\mathcal{E}_{0}^{(l)}|/c_{1}\equiv K_{0}$ since CPs with $K>K_{0}$
have $\mathcal{E}_{K}^{(l)}>0$ and thus break up \cite{schrieffer64}.

Explicit expressions of relevant thermodynamic variables and transport
coefficients evaluated within the weak-coupling limit of the $l$-wave BCS
theory have been derived in Refs.\cite{anderson61,hirschfeld93,maki94}. In
these papers it is shown that the \textit{average} behavior of most of these
quantities over the cylindrical Fermi surface exhibits small variation due
to the explicit realization of an $l=0$ or $l=2$ symmetry \cite%
{hirschfeld93,maki94}. In particular, the temperature-dependent gap equation
is given by \cite{schrieffer64,anderson61, maki94} 
\begin{equation}
1=\mathcal{N}_{0}V_{0}\int_{0}^{\hbar \omega _{D}}d\epsilon _{\mathbf{k}}{%
\Big \langle}\frac{\tanh \left( \frac{1}{2}\beta \sqrt{\epsilon _{\mathbf{k}%
}^{2}+\Delta ^{(l)^{2}}|g^{(l)}(\mathbf{k})|^{2}}\right) }{\sqrt{\epsilon _{%
\mathbf{k}}^{2}+\Delta ^{(l)^{2}}|g^{(l)}(\mathbf{k})|^{2}}}{\Big \rangle}%
_{F}  \label{gap}
\end{equation}%
where $\beta =1/k_{B}T$, $g^{(0)}(\mathbf{k})=1$ for $l=0$, and $g^{(2)}(%
\mathbf{k})=\cos (2\theta )$ for $l=2$. The critical temperature is
determined from (\ref{gap}) by the condition $\Delta ^{(l)}(T_{c})=0$. In
the weak-coupling limit $\hbar \omega _{D}/k_{B}T_{c}\gg 1$ it can be
calculated analytically and it follows that $T_{c}$ is independent of the $l-
$state \cite{anderson61}: 
\begin{equation}
k_{B}T_{c}=\frac{2e^{\gamma }}{\pi }\hbar \omega _{D}\exp (-1/\mathcal{N}%
_{0}V_{0})\simeq 1.13\hbar \omega _{D}\exp (-1/\mathcal{N}_{0}V_{0})
\label{critical0}
\end{equation}%
with $\gamma \simeq 0.577\cdots $ Euler's constant. In the zero-temperature
limit $\Delta _{0}^{(l)}\equiv \Delta ^{(l)}(T=0)$ the energy integration in
(\ref{gap}) leads to the gap relation \cite{anderson61} 
\begin{equation}
\Delta _{0}^{(l)}=2\Gamma ^{(l)}\hbar \omega _{D}\exp (-1/\mathcal{N}%
_{0}V_{0}^{(l)})  \label{gap2}
\end{equation}%
where $\Gamma ^{(l)}=\exp [-{\big \langle}|g^{(l)}|^{2}\ln |g^{(l)}|{\big
\rangle}_{F}]$. For $l=0$, $\Gamma ^{(0)}=1$, while for $l=2$, $\Gamma
^{(2)}=2\exp (-1/2)\simeq 1.213$, so that combining (\ref{critical0}) and (%
\ref{gap2}) we are led to the gap-to-$T_{c}$ ratios 
\begin{equation}
\frac{2\Delta _{0}^{(0)}}{k_{B}T_{c}}\simeq 3.53\text{ \ \ \ \ \ }\ \ \ 
\frac{2\Delta _{0}^{(2)}}{k_{B}T_{c}}\simeq 4.28.  \label{gaps}
\end{equation}%
For $l=0$ one recovers the standard BCS result \cite{BCS} and the somewhat
higher value for the $l=2$ $d$-wave case. We note that the quantity $\Delta
_{0}^{(l)}/\Gamma ^{(l)}$ has the same functional dependence as the
zero-temperature gap of the BCS theory \cite{BCS}. Considering that
measurements of the energy gap for any given cuprate show some scatter about
a central value $\Delta _{0}^{exp}$ \cite{poole95} in the following we shall
assume that $\Delta _{0}^{exp}\simeq \Delta _{0}^{(2)}/\Gamma ^{(2)}\simeq
\Delta _{0}^{(0)}\equiv \Delta _{0}$.

On the other hand, the average superfluid density $\rho _{s}(T)\equiv
\lambda _{ab}^{2}(0)/\lambda _{ab}^{2}(T)$ exhibits a more pronounced
angular momentum dependence. This is given by \cite{maki94} 
\begin{equation}
\rho _{s}^{(l)}=1-\beta \int_{0}^{\infty }d\epsilon _{\mathbf{k}}{\Big
\langle}\frac{|g^{(l)}(\mathbf{k})|^{2}}{\cosh ^{2}\left( \frac{1}{2}\beta 
\sqrt{\epsilon _{\mathbf{k}}^{2}+\Delta ^{(l)^{2}}|g^{(l)}(\mathbf{k})|^{2}}%
\right) }{\Big \rangle}_{F}.  \label{rho}
\end{equation}%
In the low-temperature limit (\ref{rho}) yields for $l=0$ an exponential $T$%
-dependence 
\begin{equation}
\rho _{s}^{(0)}(T)\simeq 1-\left( \frac{2\pi \Delta _{0}^{(0)}}{k_{B}T}%
\right) ^{1/2}\exp (-\Delta _{0}^{(0)}/k_{B}T)  \label{sf1}
\end{equation}%
while for $l=2$ it gives the linear $T$-dependence 
\begin{equation}
\rho _{s}^{(2)}(T)\simeq 1-\frac{(2\ln 2)k_{B}T}{\Delta _{0}^{(2)}}.
\label{sf2}
\end{equation}%
Experiments \cite{panagopoulos98,broun07} on the temperature variation of
the magnetic penetration depth $\lambda _{ab}(T)$ are consistent with the
quasi-linear behavior (\ref{sf2}) which is a signature of $d$-wave symmetry.
However, its asymptotic value $\lambda _{ab}(T\rightarrow 0)$ is independent
of $l$, a result that we apply below.

\section{Bose-Einstein condensation}

We assume that charge carriers are an ideal binary mixture of
non-interacting unpaired fermions plus breakable bosonic linearly-dispersive
CPs \cite{fujita02,dellano98,FG,schrieffer64}. Let the fermion number per
unit area be $n_{f}=n_{f1}+n_{f2}$ where $n_{f1}$ and $n_{f2}$ are the
number densities of unpairable and pairable fermions, respectively.
Unpairable fermions lie outside the interaction region of (\ref{interaction}%
) unlike the pairable fermions whose $T$-dependent density $n_{f2}(T)$ is%
\begin{equation}
n_{f2}(T)=2\big[\ n_{0}^{2D}(T)+n_{0<K\leq K_{0}}^{2D}(T)\ \big]%
+n_{f2}^{u}(T).  \label{fermi}
\end{equation}%
Here $n_{0}^{2D}$ is the bosonic number density of CPs with CMM wavenumber $%
K=0$, $n_{0<K<K_{0}}^{2D}$ that with $0<K<K_{0}$, and $n_{f2}^{u}$ the
number density of pairable but unpaired fermions. By asserting that in
thermal equilibrium these kinds of fermions arise precisely from broken CPs 
\cite{dellano98} we identify $n_{f2}^{u}(T)=2n_{K_{0}<K<K_{max}}^{2D}(T)$.
On the other hand, at $T=0$ all pairable fermions should belong to the
condensate (Ref.\cite{poole95}, p. 122) so that $n_{f2}(0)=2n_{0}^{2D}(0)%
\equiv 2n^{2D}$ where $n^{2D}$ is the total boson number per unit area. The
number equation for pairable fermions may thus be reexpressed in terms of
boson quantities alone, namely $n^{2D}=n_{0}^{2D}(T)+n_{0<K\leq
K_{0}}^{2D}(T)+n_{K_{0}<K\leq K_{max}}^{2D}(T)$ $\equiv
n_{0}^{2D}(T)+n_{0<K\leq K_{max}}^{2D}(T)$. Thus 
\begin{equation}
n^{2D}=n_{0}^{2D}(T)+\int_{0^{+}}^{K_{max}}\frac{d^{2}K}{(2\pi )^{2}}\frac{1%
}{z^{-1}\exp \beta \mathcal{E}_{K}^{(l)}-1}  \label{bose}
\end{equation}%
where $\beta \equiv 1/k_{B}T$, $\mu $ the boson chemical potential and $%
z\equiv \exp \beta \mu $ is the fugacity ($0\leq z\leq 1$). On introducing (%
\ref{bose}) the energy-shifted boson dispersion relation $\mathcal{E}%
_{K}^{(l)}=\hbar c_{1}K$ for $K>0$ the integral can evaluated by changing to
the variable $x\equiv \beta \hbar c_{1}K$. Since $c_{1}=2v_{F}/\pi $ and $%
K_{max}=k_{F}(1+k_{D}^{2}/k_{F}^{2})$, the upper integration limit $x_{max}$
in (\ref{bose}) is then be very large, namely $x_{max}=\beta \hbar
v_{F}k_{F}=2E_{F}/k_{B}T\gg 1$. The last inequality is consistent with the
maximum empirical value for the ratio $k_{B}T_{c}/E_{F}\leq 0.05$ reported 
\cite{uemura04}\ for all SCs including cuprate SCs. Given the rapid
convergence of Bose integrals the upper integration limit $x_{max}$\ may
safely be taken as infinite in (\ref{bose}) so that the integrals can then
be evaluated exactly by expanding the integrand in powers of $z\exp (-x)$
and integrating term by term. The number density (\ref{bose}) becomes%
\begin{equation}
n^{2D}=n_{0}^{2D}(T)+\frac{(k_{B}T)^{2}}{2\pi \hbar ^{2}c_{1}^{2}}%
\sum_{n=1}^{\infty }\frac{z^{n}}{n^{2}}  \label{density}
\end{equation}%
The critical BEC temperature $T_{c}$ is now determined by solving (\ref%
{density}) for $n_{0}^{2D}(T_{c})=0$ and $z(T_{c})=1$. One obtains 
\begin{equation}
T_{c}=\frac{\hbar c_{1}}{k_{B}}\left( \frac{2\pi n^{2D}}{\zeta (2)}\right)
^{1/2}  \label{critical1}
\end{equation}%
where $\zeta (2)=\pi ^{2}/6$.

\begin{figure}[t]
\begin{center}
\includegraphics[scale=0.45]{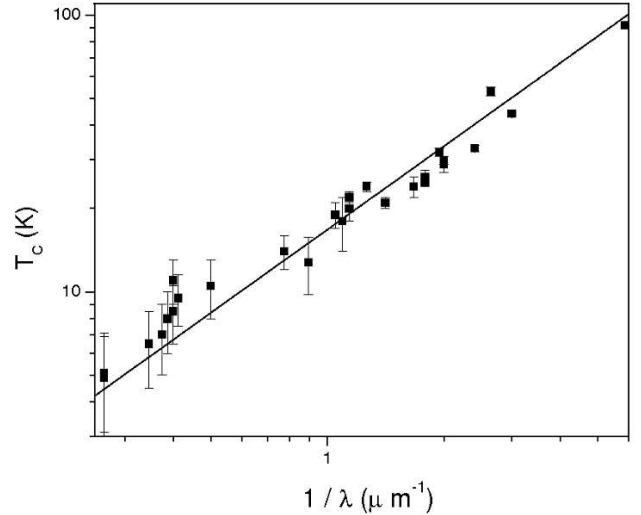}
\end{center}
\caption{Comparison of experimental $T_{c}$s vs. theoretical predictions (%
\protect\ref{curve}) as function of zero-temperature inverse penetration
length $\protect\lambda _{ab}^{-1}$ for YBCO compounds with different doping
degrees. Square datapoints are taken from Ref.\protect\cite{zuev05}, except
for uppermost square referring to the optimally doped regime \protect\cite%
{poole95}. Vertical \textquotedblleft error bars\textquotedblright\
represent full widths of $\protect\sigma _{1}$ peaks, where $\protect\sigma %
_{1}$ is the real part of the conductivity $\protect\sigma $ employed in Ref.%
\protect\cite{zuev05} to determine $\protect\lambda _{ab}^{-1}$.}
\end{figure}

\begin{table*}[tbp]
\begin{center}
\begin{tabular}{ccccccccc}
\hline\hline
superconductor & $\Theta _{D}$ (K)$^{a}$ & $\Delta _{0}$ (meV)$^{b}$ & $%
\lambda _{ab}$ (nm)$^{c}$ & $\delta (\mathring{A})$$^{d}$ & $T_{c}^{exp}$(K)$%
^{e}$ & $T_{c}^{th}$ (K) & $(2\Delta _{0}/k_{B}T_{c})^{exp}$$^{(f)}$ & $%
(2\Delta _{0}/k_{B}T_{c})^{th}$ \\ \hline
\hspace{1mm} (La$_{.925}$Sr$_{.075})_{2}$CuO$_{4}$ \hspace{1mm} & 360 & 6.5
& 250 & 4.43$^{g}$ & 36 & 36.4 & 4.3 & 4.14 \\ 
YBa$_{2}$Cu$_{3}$O$_{6.60}$ & 410 & 15.0 & 240 & 2.15$^{h}$ & 59 & 56.0 & 
5.90 & 6.09 \\ 
YBa$_{2}$Cu$_{3}$O$_{6.95}$ & 410 & 15.0 & 145 & \ 2.15$^{g}$ & 93.2 & 92.6
& 4.0 & 3.68 \\ 
Bi$_{2}$Sr$_{2}$CaCu$_{2}$O$_{8}$ & 250 & 16.0 & 250 & 2.24$^{g}$ & 80 & 72.2
& 4.64 & 4.85 \\ 
Bi$_{2}$Sr$_{2}$Ca$_{2}$Cu$_{3}$O$_{10}$ & 260 & 26.5 & 252 & 2.24$^{i}$ & 
108 & 109.2 & 5.7 & 4.99 \\ 
Tl$_{2}$Ba$_{2}$Ca$_{2}$Cu$_{2}$O$_{8}$ & 260 & 22.0 & 221 & 2.14$^{i}$ & 110
& 104.1 & 4.5 & 4.47 \\ 
\hspace{3mm} Tl$_{2}$Ba$_{2}$Ca$_{2}$Cu$_{3}$O$_{10}$ \hspace{3mm} & 280 & 
14.0 & 200 & 4.30$^{i}$ & 125 & 105.5 & 3.1 & 2.96 \\ \hline\hline
\end{tabular}%
\end{center}
\caption{ Physical parameters of cuprate superconductors and predicted
values for $T_{c}$, and the ratio $2\Delta _{0}/k_{B}T_{c}$ according to (%
\protect\ref{critical3}). Debye temperature is $\Theta _{D}\equiv \hbar 
\protect\omega _{D}/k_{B}$. Parameters taken from from Ref.\protect\cite%
{poole95} (see also references cited therein): a) table 4.1, b) table 6.1,
c) table A.1, d) table A.2, e) table A.1, f) table 6.1, g) estimated from
band-structure calculations \protect\cite{krakauer88,pickett89}, h)
estimated as $\protect\delta =0.64\ c_{int}$, and i) estimated as $\protect%
\delta =0.68\ c_{int}$, where $c_{int}$ is the CuO$_{2}$ interlayer
separation for a given cuprate. For YBCO $\Theta _{D}$, $\Delta _{0}$, and $%
\protect\delta $ are assumed the same for different dopings.}
\end{table*}

\section{Charge carrier density}

The areal density of charge carriers was formerly estimated from
measurements of the London penetration depth $\lambda _{L}$ which is the
distance over which an external magnetic field decays within the
superconductor. For superelectrons with a 3D density $n_{s}$, charge $e$,
and effective mass $m^{\ast }$, one has the well-known relation $1/\lambda
_{L}^{2}=4\pi e^{2}n_{s}/m^{\ast }c^{2}$. By introducing \cite{uemura04} the
average interlayer spacing $c_{int}$ between $CuO_{2}$ planes in HTSCs it
follows that $n^{2D}\simeq c_{int}n_{s}$. Penetration-depth data spanning a
wide range of critical temperatures are consistent with the phenomenological
Uemura relation $T_{c}\propto 1/\lambda _{L}^{2}\propto n^{2D}/m^{\ast }$ 
\cite{uemura04}.

Within the present model we evaluate the magnetic penetration depth due to
linearly-dispersive CPs with charge $2e$, and constrained to move within a
thin layer of width $\delta $ with a uniform CM speed $c_{1}$. Thus, we
first consider the expression for the 3D supercurrent of excited CPs \cite%
{fujita02} $\mathbf{J}_{s}=n^{3D}(2e)c_{1}\hat{\mathbf{K}}$ with $\hat{%
\mathbf{K}}\equiv \mathbf{K}/K$. We now introduce the contour integral of
the CP wavefunction phase within a homogeneous medium and in the presence of
an external magnetic field $\mathbf{B}=\nabla \times \mathbf{A}$. The
following integral along any closed path vanishes%
\begin{equation}
\oint \left( \hbar \mathbf{K}+\frac{2e}{c}\mathbf{A}\right) \cdot d\mathbf{r}%
=0  \label{path}
\end{equation}%
where $c$ is the speed of light in vacuum. By expressing $\mathbf{K}$ in
terms of $\mathbf{J}_{s}$ and using Stoke's theorem to evaluate (\ref{path})
we get a modified version of London's equation $\mathbf{J}_{s}=-\Lambda _{p}%
\mathbf{A}$ where $\Lambda _{p}\equiv 4e^{2}c_{1}n^{3D}/\hbar cK$. Taking
now the curl of this modified London equation and introducing Ampere's law $%
\nabla \times \mathbf{B}=(4\pi /c)\mathbf{J}_{s}$, it follows that the
magnetic induction $\mathbf{B}$ satisfies the Helmholtz equation $\nabla ^{2}%
\mathbf{B}=\lambda ^{-2}\mathbf{B}$, where%
\begin{equation}
\frac{1}{\lambda ^{2}}\equiv \frac{(2e)^{2}}{c^{2}}\frac{4\pi c_{1}n^{3D}}{%
\hbar K}.  \label{london}
\end{equation}%
Note that London's result is recovered for quasi-particles with density $%
n^{3D}\rightarrow n_{s}/2$, momentum $\hbar K\rightarrow 2m^{\ast }c_{1}$,
and charge $2e\rightarrow e$. This expression for $\lambda $ varies between
its minimum value $\lambda =0$ when $K=0$ (perfect diamagnetism), and its
maximum, say $\lambda _{0}$, for $K=K_{0}$ (CP breakup). It seems natural to
identify $\lambda _{0}$ with the experimentally observed value of the
in-plane penetration depth at $T=0$, namely $\lambda _{0}=\lambda _{ab}(T=0)$%
. Here, $\lambda _{ab}^{-1}(0)=\lambda _{a}^{-1}(0)+\lambda _{b}^{-1}(0)$ is
the geometric mean of this parameter measured along crystallographic
in-plane directions $a$ and $b$. As shown in \S\ II, this parameter is
independent of the explicit value of the angular momentum $l$. By
substituting the dispersion relation (\ref{dispersion}) to eliminate $K_{0}$
from $\lambda _{0}$ and imposing the relation $n^{2D}=\delta n^{3D}$ the 2D
charge carrier density becomes 
\begin{equation}
n^{2D}=\frac{e^{2}}{c^{2}}\frac{\delta |\mathcal{E}_{0}^{(l)}|}{16\pi
c_{1}^{2}}\frac{1}{\lambda _{ab}^{2}}.  \label{n2d}
\end{equation}%
This latter expression can be reformulated by considering the relation (\ref%
{gap2}) and the weak-coupling limit of (\ref{epsilon0}). It follows that $|%
\mathcal{E}_{0}^{(l)}|=(\Delta _{0}^{(l)})^{2}/2\hbar \omega _{D}$ so that 
\begin{equation}
n^{2D}=\frac{e^{2}}{32\pi c_{1}^{2}c^{2}}\frac{\delta \Delta _{0}^{2}}{\hbar
\omega _{D}}\frac{1}{\lambda _{ab}^{2}}  \label{n2dd}
\end{equation}%
where the approximate relation $\Delta _{0}^{(l)}/\Gamma^{(l)}\simeq \Delta
_{0}$ as justified in \S\ II was used.

\section{Critical temperature}

The final explicit expression for the critical BEC temperature $T_{c}$ is
now obtained by substituting (\ref{n2dd}) in (\ref{critical1}). This leaves 
\begin{equation}
T_{c}=\frac{\hbar c}{2\pi k_{B}e}\left( \frac{3\delta }{2\hbar \omega _{D}}%
\right) ^{1/2}\frac{\Delta _{0}}{\lambda _{ab}}  \label{critical3}
\end{equation}%
which is independent of the CP speed $c_{1}$. We observe that for fixed
values of $\omega _{D}$, $\Delta _{0}$, and $\delta $, the critical
temperature increases linearly with $\lambda _{ab}^{-1}$. This dependence
has been observed by Zuev $et$ $al.$\cite{zuev05} in experiments in
underdoped YBCO films with $T_{c}$s ranging from $6$ to $50K$. They conclude
that, within some noise, their data fall on the same curve $\rho _{s}\propto
\lambda _{ab}^{-2}\propto T_{c}^{2.3\pm 0.4}$, irrespective of annealing
procedure, oxygen content, etc. Thus, by assuming that except for $\lambda
_{ab}$ the other YBCO parameters are approximately independent of the doping
level, we introduced in (\ref{critical3}) the values: $\Theta _{D}=410K$ 
\cite{poole95}, $\Delta _{0}=14.5$ meV \cite{poole95}, and $\delta =2.15\ 
\mathring{A}$ \cite{krakauer88,pickett89} to get the relation 
\begin{equation}
T_{c}=\frac{16.79[(\mu m)^{-1}K]}{\lambda _{ab}}.  \label{curve}
\end{equation}%
Figure 1 is adapted from Ref.\cite{zuev05} and compares theoretical
predictions (\ref{curve}) with experimental data, as well as with data
pertaining to higher doping regimes. We see that (\ref{curve}) gives an
excellent fit to the experimental data. The same functional dependence has
been observed in single YBCO crystals near the optimally-doped regime \cite%
{pereg04}. More recently, Broun $et$ $al.$ \cite{broun07} found that their
samples of high-purity single-crystal YBCO followed the rule $T_{c}\propto
\lambda _{ab}^{-1}\propto n_{s}^{1/2}\propto (p-p_{c})^{1/2}$ where the
doping $p$ is the number of holes per copper atom in the $CuO_{2}$ planes and%
$\ p_{c}$ the minimal doping for superconductivity onset. The measured value
of the penetration length in YBCO crystals is an order of magnitude bigger
than in thin films \cite{broun07,pereg04}, so that the specific values of $%
T_{c}$s derived from (\ref{critical3}) are not in such good agreement as in
the YBCO films. However, one should expect variations of parameters such as
the energy gap associated to crystals and film systems. It has been pointed
out \cite{zuev05} that YBCO \textit{films} seem to behave more like other
cuprates such as BiSrCaCuO or LaSrCuO than do YBaCuO \textit{crystals}.
Furthermore, a different approach \cite{liang05} based on measurements of
the lower critical magnetic field $H_{c1}(T)$ for highly underdoped YBCO
indicates that experimental data may be consistently described only by
assuming $T_{c}\propto n_{s}^{0.61}$, in close agreement with studies
mentioned above.

Theoretical values of $T_{c}$ for superconducting cuprates with different
compositions have been also calculated using (\ref{critical3}). Here we
report on these seven layered-cuprate superconducting compounds: (La$_{.925}$%
Sr$_{.075})_{2}$CuO$_{4}$; YBa$_{2}$Cu$_{3}$O$_{6.60}$; YBa$_{2}$Cu$_{3}$O$%
_{6.95}$; Tl$_{2}$Ba$_{2}$Ca$_{2}$Cu$_{2}$O$_{8}$; Tl$_{2}$Ba$_{2}$Ca$_{2}$Cu%
$_{3}$O$_{10}$; Bi$_{2}$Sr$_{2}$Ca$_{2}$Cu$_{3}$O$_{10}$; and Bi$_{2}$Sr$%
_{2} $CaCu$_{2}$O$_{8}$. Characteristic parameters for these materials were
taken from tables compiled in Ref.\cite{poole95} (see also \cite%
{harshman92,hasegawa89,ginsberg92}). Concerning the layer width $\delta $ no
direct experimental data are available. We have employed results derived
from energy band-structure calculations for cuprates. Contour plots \cite%
{krakauer88,pickett89} of the charge distribution for La$_{2}$CuO$_{4}$, YBa$%
_{2}$Cu$_{3}$O$_{7}$, and BiCa$_{2}$SrCu$_{2}$O$_{8}$ suggest that charge
carriers in each of these systems are concentrated within slabs of average
width $\delta \simeq 2.61\mathring{A}$, $2.15\mathring{A}$, and $2.28%
\mathring{A}$, respectively, about their $CuO_{2}$ planes. As $c_{int}$
denotes the average separation between adjacent $CuO_{2}$ planes, it follows
from crystallographic data \cite{poole95} that the yttrium and bismuth
compounds give $\delta \simeq 0.64\ c_{int}$ and $0.68\ c_{int}$,
respectively. Taking into account that BiSr$_{2}$Ca$_{n}$Cu$_{n+1}$O$_{6+n}$
compounds possess the same layering scheme as their TlBa$_{2}$Ca$_{n}$Cu$%
_{n+1}$O$_{6+n}$ counterparts \cite{poole95}, we assumed that the condition $%
\delta \simeq 0.68c_{int}$ holds also for the thallium compounds. The former
estimations are congruent with Uemura's surmise \cite{uemura04} that SC
charge carriers in layered cuprates are concentrated within slabs of width $%
\delta =c_{int}$.

Table I shows results obtained using the foregoing assumptions, together
with the physical parameters involved in the calculation. In most cases we
find rather satisfactory agreement between predicted and measured values of $%
T_{c}$. We also find very good agreement between theoretical and
experimental gap-to-$T_{c}$ ratios $2\Delta _{0}/k_{B}T_{c}$. Average
theoretical and experimental such ratios presented in Table I are $(2\Delta
_{0}/k_{B}T_{c})^{th}\simeq 4.45$ and $(2\Delta _{0}/k_{B}T_{c})^{exp}\simeq
4.59$, respectively. Both are consistent with the ratio $2\Delta
_{0}^{(2)}/k_{B}T_{c}\simeq 4.28$ predicted by the $l=2$ BCS theory in (\ref%
{gaps}). We have not attempted estimate uncertainties of our theoretical
results since the accumulated data of the physical parameters involved in
the calculation, particularly $\Delta _{0}$ and $\lambda _{ab}$, show a wide
scatter.

\section{Discussion and conclusions}

We have shown that layered-cuprate HTSC can be described by means of an $l$%
-wave BCS theory for a quasi-2D BEC of Cooper pairs. The theory involves a 
\textit{linear}, as opposed to quadratic, dispersion relation in their total
or CM momenta. The theory yields a simple formula for the critical
transition temperature $T_{c}$ with a functional relation $T_{c}\propto
1/\lambda _{ab}\propto n_{s}^{1/2}$ which applies to a variety of cuprate
SCs over a wide range of dopings. Although this behavior apparently
disagrees with the phenomenological Uemura relation $T_{c}\propto 1/\lambda
_{ab}^{2}$ \cite{uemura04}, different experimental studies \cite%
{broun07,zuev05,pereg04} show consistency with the inverse linear dependence
of $T_{c}$. Additional consistency is also seen with the reported dependence 
$T_{c}\propto n_{s}^{0.61}$ arising from measurements of the lower critical
magnetic field \cite{liang05}. When averaged over a cylindrical Fermi
surface, the physical quantities involved in the theory show small
dependence on the angular momentum state $l$. However, the gap-to-$T_{c}$
ratio $2\Delta _{0}/k_{B}T_{c}$ is closer to that predicted by the extended
BCS theory for $l=2$ than for $l=0$. It is shown elsewhere \cite%
{villarreal09} that all relevant 2D expressions derived here arise in the
limit $k_{B}T\delta /\hbar c_{1}\rightarrow 0$ of a more general 3D BCS-BEC
theory for layered materials.

$\ast $Permanent address.

\textbf{Acknowledgments } We thank M. Fortes, S. Fujita, L.A. P\'{e}rez, and
M.A. Sol\'{\i}s for fruitful discussions. MdeLl thanks UNAM-DGAPA-PAPIIT
(Mexico) IN106908 as well as CONACyT (Mexico) for partial support. He thanks
D.M. Eagles and R.A. Klemm for e-correspondence and is grateful to W.C.
Stwalley for discussions and the University of Connecticut for its
hospitality while on sabbatical leave.

\end{document}